\documentclass[aps,prl,superscriptaddress,twocolumn,amsfonts,amsmath,a4paper,showpacs]{revtex4}
\usepackage{graphicx}

\begin{document}

\title{Diffusion and Relaxation Dynamics in Cluster Crystals}

\author{Angel J.\ Moreno}
\affiliation{\mbox{Institut f\"{u}r Theoretische Physik II, 
Heinrich-Heine-Universit\"{a}t D\"{u}sseldorf, 
D-40225 D\"{usseldorf}, Germany}}
\affiliation{\mbox{Centro de F\'{\i}sica de Materiales (CSIC-UPV/EHU), 
Apartado 1072, E-20080 San Sebasti\'{a}n, Spain}}

\author{Christos N.\ Likos}
\affiliation{\mbox{Institut f\"{u}r Theoretische Physik II, 
Heinrich-Heine-Universit\"{a}t D\"{u}sseldorf, 
D-40225 D\"{usseldorf}, Germany}}

\begin{abstract}

For a large class of fluids exhibiting ultrasoft bounded pair potentials, particles
form crystals consisting of clusters located in the lattice sites, with 
a density-independent lattice constant. Here we present an investigation
on the dynamic features of a representative example of this class.
It is found that particles can diffuse between lattice sites, maintaining the lattice structure,
through an activated hopping mechanism. This feature yields finite values for the diffusivity 
and full relaxation of density correlation functions. 
Simulations suggest the existence of a localization transition which is avoided by hopping,
and a dynamic decoupling between self- and collective correlations.

\end{abstract}

\date{\today}
\maketitle

The investigation of large-scale structural and dynamic 
properties of complex fluids
can be facilitated by coarse-graining intramolecular fast degrees-of-freedom. 
By following
this procedure, each macromolecule is represented 
as a single particle interacting with any other
through an effective ultrasoft pair potential \cite{likosreview}. 
For isotropic interactions, the latter
just depends on the distance between centers-of-mass. 
Details of macromolecular size and strength 
of the monomer-monomer interactions enter into 
parameters of the effective potential.
The latter is bounded 
if centers-of-mass can coincide without violating excluded volume conditions.
Some examples are polymer chains \cite{flory}, 
dendrimers \cite{gotze}, or microgels \cite{denton}.

Generalized exponential models (GEM), $v(r) = \epsilon\exp[-(r/\sigma)^m]$,
constitute a class of such effective bounded interactions. 
The cases $m \leq 2$ and $m > 2$ belong, respectively, 
to the so-called $Q^{+}$ and $Q^{\pm}$ classes
for which the Fourier transform, $\tilde{v}(q)$, of $v(r)$ is, 
respectively, positive definite or oscillating around zero.
According to a general criterion based on a mean-field density
functional theory \cite{precri}, 
systems belonging to the $Q^{+}$ class display reentrant crystallization
in the density-temperature plane \cite{lang,gottwald}. 
Potentials of the $Q^{\pm}$ class 
do not yield reentrance but rather a monotonic freezing line beyond which
the system forms cluster crystals
\cite{precri,psm,gemprl,gemlong}. 
These are novel forms for the self-organization of (soft) matter
\cite{daan:nature} in the sense that they feature a
lattice constant that is 
density-independent and the cluster population
is proportional to the density \cite{precri,gemprl,gemlong}.
Recent simulations of amphiphilic dendrimers show that the latter
indeed interact via a $Q^{\pm}$-potential \cite{mladekdendr}. 

Previous theoretical and computational investigations 
of the GEM model for $m > 2$ have focused
on structural and thermodynamic properties \cite{gemprl,gemlong}. 
The {\it dynamics} of these crystals have not been studied to-date
and their particular form of self-organization
suggests that the former must be different from that of atomic solids
with single site occupancy.
The purpose of this Letter is to investigate the novel features arising
in the dynamics of such cluster crystals in general.
We find that particles diffuse 
between lattice sites through an activated hopping mechanism, 
without breaking the lattice structure, resulting in 
finite values for the diffusivity and full relaxation 
of self-correlation functions. 
We establish the existence of a localization transition which is avoided by
hopping and is distinct for self- and density-density correlations. Therefore,
an unusual decoupling between self- and collective dynamics is observed,
the latter exhibiting a lower transition temperature than the former.

Particles in the simulated system interact through a 
GEM potential with $m = 8$,
which is cut-off at a distance $r_{\rm c} = 1.5\sigma$. 
The particle mass $m_{\rm p}$,
energy scale $\epsilon$, and particle diameter $\sigma$
are set to one. In the following, energy, temperature, density, time, distance, and wavevector 
will be given respectively in units of $\epsilon$, $\epsilon/k_{\rm B}$, $\sigma^{-3}$, $\sigma(m_{\rm p}/\epsilon)^{1/2}$,
$\sigma$, and $\sigma^{-1}$. 
We investigate dynamic features of the cluster crystals, 
in a wide range of temperature $T$,
deep inside the region of stability of the fcc-phase \cite{gemprl,gemlong}.  
The investigated densities are $\rho = N/L^3 = $ 2.0, 3.0, 4.0, and 7.0. 
The values of the cell size, $L$, used for periodic boundary conditions
are integer multiples of the $\rho$-independent lattice constant $a$. Following \cite{gemprl,gemlong}, for a fcc lattice
$a = 2\pi\sqrt{3}/q_{\ast}$, with $q_{\ast}$ the wavevector for which $\tilde{v}(q)$ shows its first minimum.
The obtained value is $a = 1.8571325$. The distance between nearest-neighbor lattice sites is
$d_{\rm nn} = a/\sqrt{2} = 1.313191$.

Due to the large values of $\rho$, investigation of relaxation dynamics
in the present system is computationally more demanding that for standard 
model systems with unbounded interactions as, e.g., Lennard Jones mixtures \cite{gleimkob}, which are 
typically studied at $\rho \sim 1.0$. For this reason 
Newtonian dynamics simulations have been carried out instead of Brownian dynamics,
which would yield a much slower relaxation \cite{notebrown}. 

\begin{figure}
\includegraphics[width=0.75\linewidth]{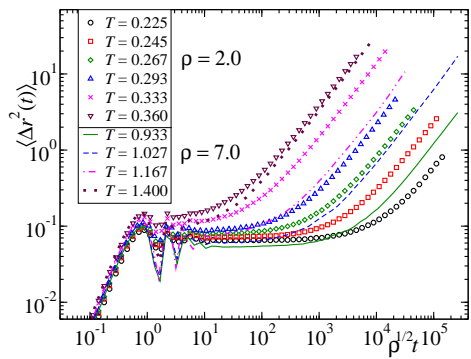}
\caption{Temperature dependence of the mean squared displacement
for density $\rho = 2.0$ (symbols) and $7.0$ (lines).}
\label{fig:msd}
\end{figure}

\begin{figure}
\includegraphics[width=0.77\linewidth]{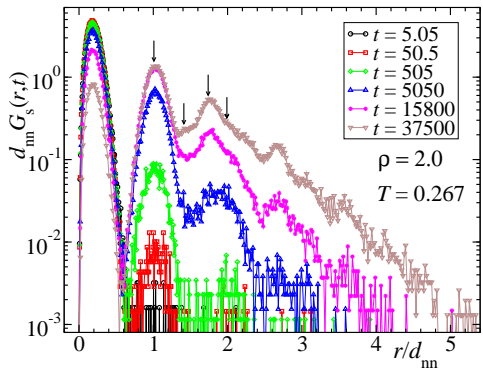}
\caption{
The van Hove self-correlation function,
$d_{\rm nn}G_{\rm s}(r,t)$ computed at different times, 
vs.\ distance $r/d_{\rm nn}$,
where $d_{\rm nn}$ is the distance between nearest-neighbor lattice sites,
at $\rho = 2.0$ and $T = 0.267$.
Arrows correspond to distances between $n$th-neighbor lattice sites, 
for $n =$ 1, 2, 3, and 4.}
\label{fig:vhove}
\end{figure}

Fig.\ \ref{fig:msd} shows the $T$-dependence of the mean squared displacement for
densities $\rho = 2.0$ and $\rho = 7.0$. The initial ballistic regime
corresponds to oscillations around lattice sites whose period scales,
according to \cite{gemlong} as $\rho^{-1/2}$. Rescaling time as
$\rho^{1/2}t$ causes a collapse of the curves at short times and confirms the
above result that was derived on the basis of 
statics \cite{gemlong}. Short-time motion
is dictated, thus, by $\rho$ alone whereas long-time diffusion by
$\rho/T$, as we will shortly demonstrate. After the ballistic regime,
a plateau arises characterizing the temporary trapping of the particles
within the clusters. At long times, the particles reach the diffusive regime
($\langle \Delta r ^2 (t)\rangle \sim t$) at all the investigated temperatures,
reaching distances of at least one particle diameter from their initial position.
The fcc structure remains stable in the whole
time window of the simulation. Therefore, particles move between
neighboring clusters,
modifying thus the clusters' initial identity but leaving unaffected their
average population $n_c$.

Fig.\ \ref{fig:vhove} displays an illustrative example  
of the time evolution of the van Hove self-correlation function, $G_{\rm s}(r,t)$.
The discrete nature of the motion between clusters centered around distinct lattice sites
is evidenced by the succession of peaks arising in $G_{\rm s}(r,t)$.
With increasing time, the height of the first peak progressively decreases and new peaks
of increasing intensity located at larger distances arise. From integration of $G_{\rm s}(r,t)$
for the longest represented $t$ it is found that more than a 80 \% of the particles
are located at distance $r > 0.5d_{\rm nn}$ from their initial position, i.e., they have moved to
different lattice sites. 

Though a detailed characterization of jumps between lattice sites is beyond the scope of this Letter,
it is worth commenting some interesting features of $G_{\rm s}(r,t)$.
Arrows in Fig.\ \ref{fig:vhove} indicate distances between $n$th
nearest-neighbor lattice sites, 
$\sqrt{n}d_{\rm nn}$, where $n=$ 1, 2, 3, and 4. The two first sharp
maxima after the initial peak match such distances for $n =$ 1 and 3. 
The pronounced minimum at 
$r \cong  1.4d_{\rm nn}$ matches the case $n = 2$.
Given two consecutive jumps connecting three {\it distinct} 
lattice sites 1, 2, 3 for which $d_{12} = d_{23} = d_{\rm nn}$, it is easy to see 
that the four allowed angles between both jumps,
$60^{\rm o}$, $90^{\rm o}$, $120^{\rm o}$, and $180^{\rm o}$ 
yield, respectively, $d_{13}/d_{\rm nn}=$ 
1, $\sqrt{2}$, $\sqrt{3}$, and 2. The presence of a marked minimum
for $r/d_{\rm nn} = \sqrt{2}$ and a sharp maximum for 
$r/d_{\rm nn} = \sqrt{3}$ suggests
a preferential directionality for the motion between neighboring sites with
low and high probability for angles of, respectively,  
$90^{\rm o}$ and $120^{\rm o}$ between consecutive jumps.

\begin{figure}
\includegraphics[width=0.75\linewidth]{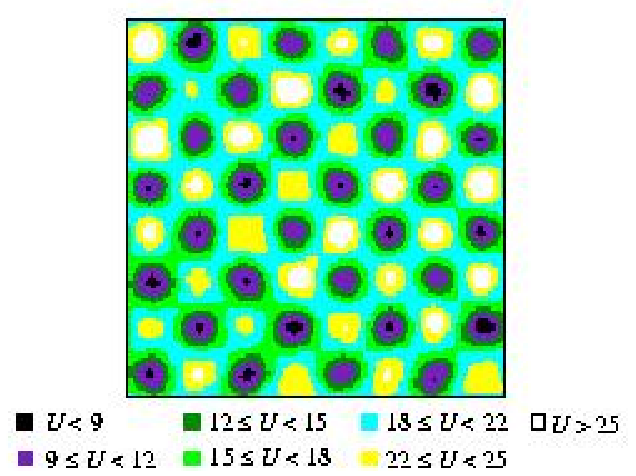}
\caption{Projection of the potential energy $U$ on a [001] lattice plane, 
for a configuration
at $\rho = 5.0$ and $T = 0.667$. Darker colors represent lower 
energies (see legend) and are centered around lattice sites.}
\label{fig:pel}
\end{figure}

\begin{figure}
\includegraphics[width=0.78\linewidth,clip]{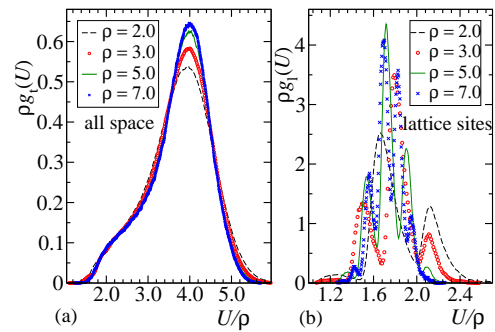}
\caption{The distribution 
of local potential energy, $\rho g(U)$,
vs.\ rescaled energy, $U/\rho$, at different densities.
(a): results considering
all the volume in the simulation box; (b): results by computing $U$ 
only at lattice sites.}
\label{fig:distener}
\end{figure}

Fig.\ \ref{fig:pel} shows for a typical configuration at $\rho = 5.0$ and $T = 0.667$
a map of the local potential energy $U$ at a [001] lattice plane,
created by {\it all} the particles in the system. 
According to Fragner \cite{fragner}, for a sufficiently small lattice constant
the superposition of potentials of the $Q^{\pm}$ class
centered at crystal lattice sites exhibits potential minima and maxima
at respectively the lattice and interstitial sites.
Results in Fig.\ \ref{fig:pel} confirm this prediction.
The potential barriers between nearest-neighbor minima (separation $d_{\rm nn}$)
are much lower than between next-nearest ones (separation $a$), suggesting
a much lower probability for single jumps of length $a$. This feature is consistent
with the minimum at $\sqrt{2}d_{\rm nn} = a$ observed in $G_{\rm s}(r,t)$ for all
the densities (see Fig.\ \ref{fig:vhove})
 
Fig.\ \ref{fig:distener} shows the distribution of local $U$-values, 
computed over the whole simulation box
[$g_{\rm t}(U)$, panel(a)] and by only considering lattice sites [$g_{\rm l}(U)$, panel(b)]. 
The distribution $g_{\rm l}(U)$ is dominated by the contribution of the individual cluster centered
around the considered site. Since the number of particles forming the cluster
is polydisperse and obviously discrete, $g_{\rm l}(U)$ shows a peak structure.
Consistently with the common value of the lattice constant and the relation $n_{\rm c} \propto \rho$,
approximate $\rho$-scaling behavior is observed for $g_{\rm t}(U)$, of increasing quality at higher densities.
The maximum at $U_{\rm t} \cong 4\rho$ corresponds to typical values at interstitials 
between nearest-neighbor sites (see Fig.\ \ref{fig:pel}).
Though the most probable $U$ for $g_{\rm l}(U)$ shows a weak $\rho$-dependence, we take
a `common' value $U_{\rm l} \cong 1.7\rho$ as a rough estimation of the average local minimum at the lattice sites.
Hence, the typical potential barrier separating nearest-neighbor lattice sites is 
$\Delta U \equiv U_{\rm t} - U_{\rm l} \cong 2.3\rho$.

\begin{figure}
\includegraphics[width=0.64\linewidth]{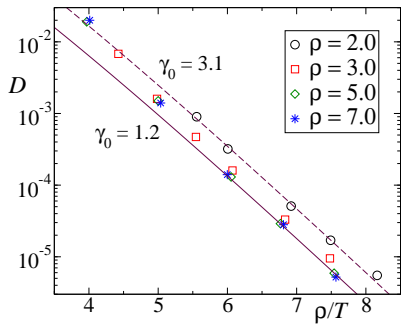}
\caption{Temperature dependence of the diffusivity for different 
densities. Lines are fits to Eq.\ (\ref{diffuse:eq}); 
the values of the fit parameter
$\gamma_0$ are indicated on the plot.}
\label{fig:diff}
\end{figure}

Fig.\ \ref{fig:diff} displays results for the diffusivity $D$, 
obtained as the long-time limit of
$\langle \Delta r^2(t)\rangle/6t$. The dependence of $D$ on density and
temperature can be understood as follows. Each particle located on
any given lattice site experiences a harmonic potential with
a barrier height $\Delta U$ to the neighboring site, as mentioned above. 
Due to the contact with the heat bath, the energy $E$ of a particle
follows the distribution $f(E) = e^{-\beta E}(\beta^3E^2)/2$
for $E > 0$ and vanishing for $E < 0$, with $\beta = (k_{\rm B} T)^{-1}$. The probability 
$P_{>}(\Delta U)$ to have $E > \Delta U$ can be calculated as
$P_{>}(\Delta U) = e^{-\beta\Delta U}[
(\beta\Delta U)^2/2 + \beta\Delta U + 1]$.
Once $E > \Delta U$, the particle hops to the neighboring site. 
We can now translate $P_{>}(\Delta U)$ into a `waiting time'
$t_w$ that particles typically spend on a given lattice site 
before hopping to the next one, which scales as 
$t_w \sim 1/P_{>}(\Delta U)$. If we observe any given particle for
a number of timesteps $N_{\rm s} \gg 1$, there will only be a fraction
$N_{\rm hop} \sim P_{>}(\Delta U) N_{\rm s}$ of them for which the
particle will hop, its motion being a random walk of
step size $\ell \sim \sigma$. Hence, measured in the
natural units of the problem, $\langle\Delta r^2(t)\rangle$ will
scale as $P_{>}(\Delta U)t$. By using  $\Delta U \cong 2.3\rho$ (see above)
we obtain
\begin{equation}
D = \gamma_0\left[(2.3\rho/T)^2/2 + 2.3\rho/T + 1\right]e^{-2.3\rho/T},
\label{diffuse:eq}
\end{equation}
with some numerical coefficient $\gamma_0$ of order unity. Theory predicts, thus,
that $D$ depends solely on the ratio $\rho/T$. The approximations involved
in deriving Eq.\ (\ref{diffuse:eq}) above become more accurate as $\rho$ grows, for which
case the polydispersity in the cluster population is reduced and all sites can
be treated as identical harmonic wells. The results shown in Fig.\ \ref{fig:diff}
fully corroborate this treatment. The diffusivity can be very well described
by the law of Eq.\ (\ref{diffuse:eq}) above and, indeed, for sufficiently high
densities all data points collapse on a single curve, where $D$ is 
expressed as a function of the ratio $\rho/T$. This is a dynamical generalization
of the scaling properties previously found for statics
\cite{gemlong}.

\begin{figure}
\includegraphics[width=0.75\linewidth]{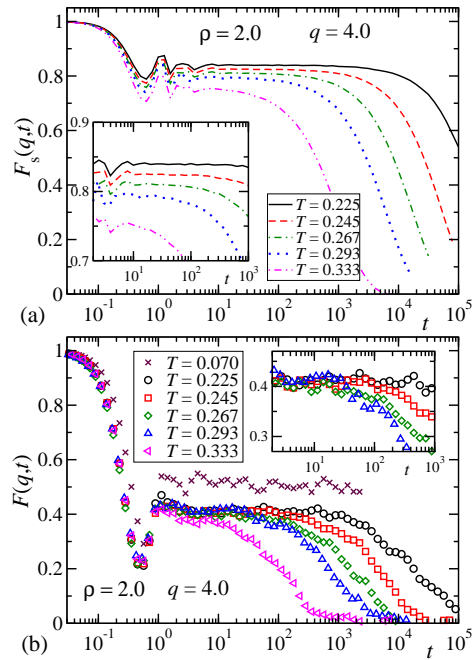}
\caption{Temperature dependence
of $F_{\rm s}(q,t)$ [(a)] and $F(q,t)$ [(b)],
for $\rho = 2.0$ and $q = 4.0$.
The insets show a magnification of the plateau regime.}
\label{fig:fqtfsqt}
\end{figure}

Finally, we discuss the self- and density-density 
correlators of wavevector $q$, respectively defined as
$F_{\rm s}(q,t) =  N^{-1}\langle \sum_{j} 
\exp[{\rm i}{\bf q}\cdot ({\bf r}_{j}(t)-{\bf r}_{j}(0))]\rangle$,
and $F(q,t) = \langle \rho_q(t) \rho^{*}_{q}(0)\rangle/
\langle \rho_q(0) \rho^{*}_q(0)\rangle$,
where $\rho_q(t)$ is defined as
$\rho_q(t) = \sum_{j}\exp[{\rm i}{\bf q}\cdot {\bf r}_j(t)]$.
Fig.\ \ref{fig:fqtfsqt} shows, 
for $\rho = 2.0$ and for a fixed wavevector $q = 4.0$
not probing the lattice structure,
the $T$-dependence of $F_{\rm s} (q,t)$ and $F(q,t)$. 
After the first microscopic decay, both correlators exhibit a plateau, whose duration
increases with decreasing $T$. As usually found in systems
displaying slow relaxation, the presence of this plateau
indicates a temporary freezing of such correlations. Strong density oscillations related with
intracluster motion are observed at the crossover between the microscopic
and plateau regimes \cite{noteosc}.
At long times self-correlations relax and decay to zero. This is also true for 
density-density correlations if $q$
does not match any wavevector probing the lattice structure. 
On the contrary, and consistently with the observed stability of the fcc lattice, 
density-density correlations for $q$ 
belonging to the reciprocal lattice are permanently frozen, and
there is no signature of a final decay of the plateau,
which remains flat (not shown) up to the limit of the simulation window.

\begin{figure}
\includegraphics[width=0.71\linewidth]{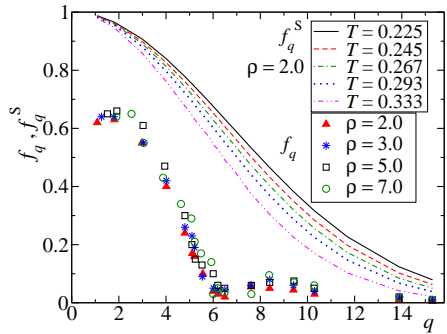}
\caption{$q$-dependence of the non-ergodicity parameters for 
self- ($f_q^{\rm s}$, lines) and density-density ($f_q$, symbols) correlators. Data for $f_q^{\rm s}$ are
displayed for fixed $\rho = 2.0$ and several temperatures. Data for $f_q$ 
are shown for different densities and correspond to state points above the localization 
transition. Data for $f_q$ are only shown for $q$-values not belonging to the reciprocal lattice.}
\label{fig:fqfs}
\end{figure}

An {\it ideal} (i.e., without intervening hopping events) localization transition 
to a non-ergodic (`glassy') phase can be defined by a jump from zero to a finite value
(non-ergodicity parameter) of the long-time limit of dynamic 
correlators \cite{mctrev}. 
Hence, at the transition point the non-ergodicity parameter is equal to the plateau height.
This quantity progressively increases with decreasing temperature in the non-ergodic phase.
The presence of hopping events can, in principle, restore ergodicity at temperatures below the ideal transition,
leading to final relaxation of correlators \cite{mctrev}. Still, the existence of an avoided ideal transition 
can still be identified by the increase of the plateau height. 
Now we investigate features of such a transition in the present system.
As shown in Fig.\ \ref{fig:fqtfsqt}, $F_{\rm s}(q,t)$ 
and $F(q,t)$ exhibit a rather
different behaviour at the intermediate plateau regime.
The plateau height, $f_q^{\rm s}$, for $F_{\rm s} (q,t)$ increases with decreasing $T$ [see inset in panel (a)]. 
This feature is observed for all wavevectors and suggests the existence of a localization transition
for self-motions. Fig.\ \ref{fig:fqfs} shows the $q$-dependence of  $f_q^{\rm s}$
for $\rho = 2.0$ at the same temperatures 
of Fig.\ \ref{fig:fqtfsqt}. The increasing width of $f_q^{\rm s}$
with decreasing $T$ indicates a progressive decrease of the 
localization length \cite{mctrev}.
However, in contrast to the standard behavior in crystal states, and similarly to relaxation in
supercooled fluids, hopping events move the particles beyond the localization length
and ergodicity for self-motions is restored, yielding 
final relaxation of $F_{\rm s}(q,t)$.

The inset in Fig.\ \ref{fig:fqtfsqt}(b) shows that, in the same $T$-range
of the latter data, the plateau height, $f_q$, for $F(q,t)$
stays constant, contrary to the case of $f_q^{\rm s}$. This behavior is observed
for all $q$-values, except for those probing the lattice structure, for which a progressive
increase of the plateau is observed with decreasing $T$ (not shown). 
An increase of $f_q$ for $q$ not probing the lattice 
structure is only observed at much lower $T$
[see data for $T = 0.070$ in Fig.\ \ref{fig:fqtfsqt}(b)]. 
Though for these low-$T$ values the system 
cannot be equilibrated and aging effects are observed at long times (not shown), such effects do not affect
the behavior of $F(q,t)$ at intermediate times, where the plateau regime arises.
Results in Fig.\ \ref{fig:fqtfsqt}(b) show that the localization transition
occurs at lower $T$ for out-of-lattice collective correlations than for self-correlations, i.e., there is
a dynamic decoupling between self- and collective relaxation.
These results resemble dynamic features of plastic crystals,
where molecules are constrained to vibrate around lattice positions, but can perform full rotations
leading to relaxation of out-of-lattice collective correlations (in the present case
cluster deformation is an additional mechanism). However, contrary to the case of plastic crystals,
activated hopping restores ergodicity of translational motions and leads to finite diffusivity
and full decay of self-correlators. 

Results for $f_q$ (excluding $q$-values probing the lattice structure)
above its localization transition (i.e., $T$-independent) are shown in 
Fig.\ \ref{fig:fqfs}
for the investigated densities. The approximate scaling behavior of $f_q$ is consistent with a common
lattice constant for all the densities. 
Since $n_{\rm c} \propto \rho$,
the major effect of increasing $n_{\rm c}$ on $F(q,t)$ is a rescaling
of $\rho_q$, which is canceled after normalization of $F(q,t)$. The smaller width of $f_q$ as compared to $f^{\rm s}_{q}$
indicates a much weaker localization for collective than for self-motions.

In summary, we have investigated slow dynamics in the cluster crystal phase of a representative model
of macromolecules in solution interacting through effective ultrasoft bounded potentials. 
The obtained major features are a full change
of the initial identity of the clusters through particle hopping between lattice sites, 
and dynamic decoupling between self- and collective out-of-lattice correlations.

We thank B. M. Mladek, P. Charbonneau, H. Fragner, and D. Frenkel for useful discussions. 
Financial support of SoftComp (NMP3-CT-2004-502235) and DIPC-Spain is acknowledged.


\begin{thebibliography}{9}
\bibitem{notebrown} Though Brownian dynamics provide a more realistic
picture at short times,  Newtonian and  Brownian dynamics are equivalent 
within a time shift at long time scales probing slow relaxation \cite{gleimkob}.
\bibitem{noteosc} 
In principle the inclusion of solvent effects through  Brownian dynamics would
damp these oscillations.
\end{thebibliography}
\end{document}